\documentstyle[12pt]{article}

\begin{document}
\bibliographystyle{unsrt}

\begin{flushright} hep-ph/9506354

BA-95-15

UMD-PP-95-117

April 1995
\end{flushright}

\vspace{6mm}

\begin{center}

{\Large \bf New Vector-Scalar Contributions to
Neutrinoless
Double Beta decay and Constraints on R-Parity Violation}\\ [6mm]

\vspace{6mm}
{\bf{K.S. Babu\footnote{Address starting September 1995: School of
Natural Sciences, Institute for Advanced Study, Olden Lane, Princeton,
NJ 08540; work supported by the Department of Energy
Grant \#DE-FG02-91ER406267}}}

{\it{Bartol Research Institute}}\\
{\it{University of Delaware}}

{\it{Newark, DE 19716}}

\vspace{3mm}

and

\vspace{3mm}

{\bf{R.N. Mohapatra\footnote{Work supported by the
 National Science Foundation Grant \#PHY-9119745}}}

{\it{ Department of Physics and Astronomy}}\\
{\it{University of Maryland}}\\

{\it{ College Park, MD 20742 }}

\end{center}

\vspace{4mm}

\begin{center}
{\bf Abstract}
\end{center}
\vspace{1mm}

We show that in the minimal supersymmetric standard model (MSSM) with
R--parity breaking as well as in the left--right symmetric model, there
are new observable contributions to neutrinoless double beta decay
($\beta\beta_{0\nu}$) arising from hitherto overlooked
diagrams involving the exchange of one W boson and one scalar boson.
In particular, in the case of MSSM, the present experimental bounds on
$\beta\beta_{0\nu}$ lifetime improves the limits on certain R--parity
violating couplings by about two orders of magnitude.  It is shown that
similar diagrams also lead to enhanced rates for $\mu^- \rightarrow e^+$
conversion in nuclei, which are in the range accessible to ongoing
experiments.

\newpage

With the standard model of electroweak interactions brilliantly confirmed
by a host of experiments, the search is on for the next level of physics at TeV
or higher scales. There exist many interesting scenarios which
address the various naturalness problems of the standard
model framework. A generic feature of many of these scenarios is that
conservation laws present in the standard model no longer remain valid.
A typical conservation law that breaks
down is the one corresponding to lepton-number symmetry.
To make any theoretical headway, one needs to know the strength of the
lepton number non--conserving interactions as well as of any
other kind of interactions that may accompany them.

It has been known for a long time [1] that neutrinoless double beta decay
($\beta\beta_{0\nu}$) is a very sensitive probe of lepton number violating
terms
in the Lagrangian such as
the Majorana mass of the light neutrinos [2], right--handed weak couplings
involving heavy Majorana neutrinos [3,4], as well as Higgs [5] and other
interactions such as those involving R--parity breaking in the supersymmetric
model [6].  The reason why this observation is interesting
is that, the steadily improving experimental limits [7] on
$\beta\beta_{0\nu}$ life-time can then be translated into more stringent
limits [8] on the parameters of these new physics scenarios.
This is an extremely valuable information to have in our search
for physics beyond the standard model.

It is the goal of this letter to point out another class of
hitherto unnoticed contributions
to $\beta\beta_{0\nu}$ decay in the following
two classes of theories: (i) Minimal supersymmetric models (MSSM) with R-parity
violation; (ii) Left-right symmetric models with a low mass $W_R$. These
new contributions are of vector-scalar type in that they involve the exchange
of a $W_L$ together with a charged scalar boson with a virtual light neutrino.
These contributions do not involve a helicity flip of the
internal light neutrino, and therefore, their amplitudes are enhanced relative
to the ordinary contribution proportional to the
light Majorana neutrino mass by roughly a
factor of $p_F/m_\nu \sim (100~MeV/1~eV) \sim 10^8$ ($p_F$ is
the Fermi momentum of the nucleons in the nuclei).  Turning
the argument around, the strength of such lepton number violating scalar boson
interactions can be constrained to be $G_{eff} \le 10^{-8}G_F$ from the
$(\beta\beta)_{0\nu}$ process (modulo nuclear matrix element uncertainties),
which indeed is a severe constraint.
The effect turns out to be most dramatic on certain R-violating
couplings in the MSSM  in that it directly involves the
R-violating couplings and superpartner masses.
In the left-right models, although
such contributions involve
unknown Higgs boson mixings, they do lead to interesting bounds for
certain range of values for these parameters.
 We also point out that similar
enhanced  vector-scalar contributions exist in other rare processes such
as $\mu^-\rightarrow e^+$ conversion in nuclei which turn out to be in the
experimentally accessible range.

The new contributions arise from the combination of two effective
four-Fermi interactions of the following type which as we will show later
can arise in several interesting classes of gauge models:
\begin{eqnarray}
H_{eff}&=&{{G_F}\over{\sqrt{2}}}(
\overline{e}\gamma_\mu(1-\gamma_5)\nu_e \overline{u}
\gamma^{\mu}(1-\gamma_5)d +\epsilon_1^{ee} \overline{d}(1-\gamma_5)u
\nu_e^TC^{-1}(1-\gamma_5)e~+ \nonumber \\
&~&\epsilon_2^{ee}\overline{d}(1-\gamma_5)\nu_e u^TC^{-1}
(1-\gamma_5)e~)~+~h.c.
\end{eqnarray}
In the above, the first term is the usual (V-A) interaction, the other two
are effective lepton number violating terms.
In order to evaluate the matrix elements between nuclear states,
we need to do Fierz reordering of the $\epsilon_2^{ee}$ term, which casts it in
the form:
$${{G_F}\over{2\sqrt{2}}}\epsilon_2^{ee}\left(\overline{d}(1-\gamma_5)u
\overline{e^c}(1-\gamma_5)\nu_e~+~{{1}\over{2}}\overline{d}\sigma^{\mu\alpha}
(1-\gamma_5)u\overline{e^c}(1-\gamma_5)\sigma_{\mu\alpha}\nu_e\right)!.$$

\noindent The
parameters $\epsilon_i^{ee}$ ($i=1,2$) characterize the new interactions that
arise in a gauge model so that any limit on them translates into
limits on the parameters on the theory leading to this interaction.

It is easy to see that the two new interaction terms in Eq. (1)
give contributions to $\beta\beta_{0\nu}$ decay which do not
depend on the neutrino mass and involve a vector current at one hadronic
vertex and a scalar current in the other (hence the name vector-scalar;
of course as just mentioned, the $\epsilon_2^{ee}$--type
scalar interaction of Eq. (1)
after Fierz reordering generates also a tensor coupling).
The resulting
effective $\Delta L=2$ Hamiltonian is then given
in momentum space by:
\begin{eqnarray}
H_{\Delta L=2}&=&~{{G^2_F}}((\epsilon_1^{ee}~+~{{\epsilon_2^{ee}}\over{2}})
\overline{u}(1+\gamma_5)d
\overline{u}\gamma_\mu(1-\gamma_5)d\overline{e}\gamma^\mu(1-\gamma_5)
{{1}\over{\gamma\cdot q}}C\overline{e}^T~\nonumber \\
&+&~{{\epsilon_2^{ee}}\over{4}}
\overline{u}\sigma^{\lambda\alpha}(1+\gamma_5)d\overline{u}\gamma_{\mu}
(1-\gamma_5)d\overline{e}\gamma^{\mu}\sigma_{\lambda\alpha}
(1-\gamma_5){{1}\over{\gamma
\cdot q}}C\overline{e}^T) + h.c.
\end{eqnarray}

\noindent In Eq. (2),
$q$ refers to the momentum of the internal light neutrino propagator.
This effective Hamiltonian of course has to be evaluated between nuclear
states. This gives rise to an effective neutrino potential as in the case
of the neutrino mass contribution to $\beta\beta_{0\nu}$ decay with the
difference that $\langle m_{\nu}/{q^2}\rangle_{Nucl.}$
gets replaced by $\langle \gamma\cdot q
/{q^2}\rangle_{Nucl.}$ (and a different operator for the tensor coupling).
To the best of our knowledge, such nuclear
matrix elements for $\beta\beta_{0\nu}$ decay have not been evaluated
in the literature. We therefore resort to a crude estimate and
assume an average value of $q$ to be equal to the Fermi momentum
$p_F$ of the nucleons in the nucleus ($\approx 100$ MeV). The present
upper limits on $m_\nu$ of about $1$ eV then translates to an upper limit
on the new interaction parameter $\epsilon$ as follows:
\begin{eqnarray}
\epsilon_{1,2}^{ee}\leq 1\times 10^{-8}.
\end{eqnarray}

We realize that due to the crudeness of our estimate of the nuclear
matrix element, the above upper limit is likely to be uncertain by
perhaps a factor of five or so. Nevertheless it is important to note that
the limit on $\epsilon_{1,2}$, barring unforeseen nuclear suppressions,
is rather
stringent and will imply important restrictions on the parameters of the
gauge models leading to Eq. (1). Let us therefore, proceed to the kind of
gauge models where the last two terms in Eq. (1) can arise at low energies.

\bigskip

\noindent{\it MSSM with R-parity violation:}

\bigskip

As is well-known, the minimal supersymmetric standard model can have
explicit [9] violation of the R-symmetry
(defined by $(-1)^{3B+L+2S}$), leading to lepton
number violating interactions in the low energy  Lagrangian.
The three possible types of couplings in the
superpotential are [10]:

\begin{eqnarray}
W^{\prime}~=~\lambda_{ijk}L_iL_jE^c_k+\lambda^{\prime}_{ijk}L_iQ_jD^c_k
+\lambda^{''}_{ijk}U^c_iD^c_jD^c_k~.
\end{eqnarray}
Here $L,Q$ stand for the lepton and quark doublet superfields, $E^c$ for
the lepton singlet superfield and $U^c,D^c$ for the quark singlet superfields.
$i,j,k$ are the generation indices and we have $\lambda_{ijk} =
-\lambda_{jik}$, $\lambda^\prime_{ijk}=-\lambda^\prime_{ikj}$.  The $SU(2)$
and color indices in Eq. (4) are contracted as follows: $L_iQ_jD_k^c =
(\nu_id_j^\alpha-e_iu_j^\alpha)D_{k\alpha}^c$, etc.  The simultaneous presence
of all three terms in Eq. (4) will imply rapid proton decay, which can be
avoided by setting the $\lambda^{\prime \prime} =0$.  In this case,
baryon number remains an unbroken symmetry while
lepton number is violated [11].

There are two types of vector--scalar contributions to $\beta\beta_{0\nu}$
and related $\Delta L = 2$ processes.  These are shown in Figures 1 and 2.  The
dominant contribution to $\beta\beta_{0\nu}$ in this model arises from Fig. 1,
where the exchanged scalar particles are the $\tilde{b}-\tilde{b}^c$ pair.
This leads to a contribution to $\epsilon_2^{ee}$ given by
\begin{eqnarray}
\epsilon_2^{ee}~\simeq~\left({{(\lambda^{\prime}_{113}}\lambda^{\prime}_{131})
\over{2\sqrt{2}G_F M^2_{\tilde{b}}}}
\right)\left({{ m_b}\over{M^2_{\tilde{b}^c}}}\right)\left(\mu {\rm tan}
\beta+A_bm_0\right)~.
\end{eqnarray}

\noindent Here $A,m_0$ are supersymmetry breaking parameters, while $\mu$ is
the
supersymmetric mass of the Higgs bosons.  tan$\beta$ is the ratio of the two
Higgs vacuum expectation values and lies in the range $1 \le {\rm tan}\beta
\le m_t/m_b \approx 60$.
For the choice of all squark masses as well as $\mu$ and the SUSY
breaking mass parameters
being of order of 100 GeV, $A=1$, tan$\beta=1$, we find from Eq. (5)
that $\lambda^{\prime}_{113}\lambda^{\prime}_{131}\leq 3\times 10^{-8}$,
which is more stringent limit on this parameter than the existing ones [12].
The present limits on these parameters are
$\lambda_{113}' \le 0.03, \lambda_{131}' \le 0.26$,
which shows that the bound derived here  from $\beta\beta_{0\nu}$ is about
five orders of magnitude more stringent on the product $\lambda_{113}'
\lambda_{131}'$.  If the exchanged scalar particles in Fig. 1 are the
$\tilde{s}-\tilde{s}^c$ pair, $b$ gets replaced by $s$ in Eq.(5) and
one obtains a limit $\lambda_{121}'\lambda_{112}
\le 1 \times 10^{-6}$, which also is  more stringent by about four orders
of magnitude
than existing limits
($\lambda_{121}' \le 0.26, \lambda_{112}' \le 0.03$).  We note that the
gluino exchange diagram [6,8]
discussed in the context of $\beta\beta_{0\nu}$ only
constrains the parameter $\lambda_{111}'$, while the vector--scalar
exchange graphs constrain several other couplings.

The diagram in Fig. 2, due to the antisymmetry of $\lambda_{ijk}$, does
not contribute to $\beta\beta_{0\nu}$ decay, but will be important for
$\mu^- \rightarrow e^+$ conversion (see discussions below).

Let us also note that there exist indirect limits on the $\lambda$
and $\lambda'$ couplings arising from the induced neutrino masses.  The
magnitudes of these masses  are given  generically by $m^\nu_{ij} \sim
\lambda_{ikl}'\lambda_{jkl}'m_km_l/(16\pi^2M_{\tilde{q}})$,
where $m_{k,l}$ stand
for the masses of the $k$th and $l$th down quarks (with a similar
expression for the $\lambda$ couplings).  The couplings
$\lambda_{i33}'\lambda_{j33} \le 6 \times 10^{-5}$
is the most severely constrained
(being proportional to the $b$--quark mass-squared) where the induced neutrino
mass has been assumed to be $\le 100~eV$, which is cosmologically safe.
The corresponding limit on
$\lambda_{131}'\lambda_{113}' \le .24$ (using $m_\nu \le 1~eV$)
is a trivial constraint to be compared
with the $\beta\beta_{0\nu}$ decay limit derived here.

\bigskip

\noindent{\it Constraints on the Left-right symmetric model:}

\bigskip

Let us consider
the minimal left-right symmetric model with
a see-saw mechanism for neutrino masses [13].  The gauge group of the model is
$SU(3)_C \times SU(2)_L \times SU(2)_R \times U(1)_{B-L}$.
The Higgs sector of the model consists of the bi-doublet field
$\phi \equiv (1/2, ~ 1/2, ~ 0)$ and triplet Higgs fields:
${\Delta_L ( 1, ~0, ~ +2 ) \oplus \Delta_R (0, ~ 1, ~ +2 ).}$
The Yukawa couplings
which are invariant under gauge and parity symmetry can be written as:
\begin{eqnarray}
{\cal L}_Y
&=&  {\overline \Psi_L} h^{\ell} \phi \Psi_R + {\overline \Psi_L}
{\tilde h}^{\ell}
 {\tilde \phi} \Psi_R  +
 \overline{Q}_L\phi h^{q}Q_R + \overline{Q}_L
{\tilde h}^{q}{\tilde {\phi}}Q_R +\nonumber \\
   & ~ & \Psi^T_L \tau_2 {\vec \tau} \cdot f {\vec \Delta_L} C^{-1} \Psi_L
                                + L\rightarrow R  + h.c.     ~~~~
\end{eqnarray}
\noindent where
$h, ~{\tilde h}$ are hermitian matrices while
 $f $ is a symmetric matrix in the generation space.  $\Psi$ and $Q$
here denote the
leptonic and quark doublets respectively.

The gauge symmetry is spontaneously broken by the vacuum expectation
values:
${\langle {\Delta_R^0} \rangle = V_R
{}~~; }$~ ${\langle \Delta_L^0 \rangle \simeq  0 ~~}; $ and
${\langle \phi \rangle = diag. ( \kappa, ~\kappa^\prime).}$
As usual, $\langle \phi\rangle $ gives masses
 to the charged fermions and Dirac masses to the neutrinos whereas
$\langle \Delta_R^0 \rangle$
 leads to the see-saw mechanism for the neutrinos in the standard way [13].

The physics we are interested in comes from the left-handed triplet
sector of the theory through its mixing with the bidoublet field
which arise from the
couplings in the Higgs potential, such as Tr$(\Delta_L \phi \Delta_R^{\dagger}
  \phi^{\dagger})$ after the full gauge symmetry is broken down to $U(1)_{em}$.
Specifically, there is a mixing between the singly charged components of
$\phi$ and $\Delta_L$ [14] (we denote this mixing term by an angle $\theta$).
This will contribute to the four-Fermi interaction
of the form given by the $\epsilon_1^{ee}$ term through the diagram shown in
Fig. 3 with
\begin{equation}
\epsilon_1^{ee}\simeq
{{h_u f_{11} sin 2\theta}\over {4\sqrt{2}G_F M^2_{H^{+}}}}~,
\end{equation}
where we have assumed
that $H^+$ is the lighter of the two Higgs fields.  We get
$h_u f_{11}{\rm sin}2\theta
\leq 6\times 10^{-9}(M_{H^+}/ 100~GeV)^2$, which is quite
a stringent constraint on the parameters of the theory. To appreciate
this somewhat more, we point out that one expects $h_u\approx m_u/ m_W
\approx 5 \times 10^{-5}$ in which case, we get an upper limit for the coupling
of the Higgs triplets to leptons $f_{11}{\rm sin}2\theta
\leq 10^{-4}$ (for $m_{H^+} = 100 ~GeV$).
Taking a reasonable choice of $\theta \sim M_{W_L}/M_{W_R}
\sim 10^{-1}$ would correspond to a limit $f_{11} \le 10^{-3}$.
Limits on this
parameters from analysis [15] of Bhabha scattering is only of order $.2$ or
so for the same value of the Higgs mass.

\vspace{3mm}

\noindent{\it $\mu^{-}\rightarrow e^{+}$ conversion:}

\vspace{3mm}

Another class of rare processes where the new vector-scalar contribution
makes an important impact is the process of $\mu^{-}\rightarrow e^+$
conversion in nuclei which arises with an observable strength in the
R violating MSSM.  This involves the couplings $\lambda_{ijk}$
which were
not constrained by the considerations of neutrinoless double beta decay
due to anti-symmetry of the Yukawa couplings.  Another contribution involves
a different product of $\lambda'$ couplings than what appeared in the
$\beta\beta_{0\nu}$ process.  We parametrize the effective four Fermi
interaction for this process in analogy to Eq. (1) with $\epsilon_1^{ee}$
replaced by two terms $\epsilon_1^{e\mu}$ and $\epsilon_1^{\mu e}$ and
similarly for $\epsilon_2$.  Here $\epsilon_1^{e\mu}$ is the coefficient of
the term $G_F/\sqrt{2}[\overline{d}(1-\gamma_5)u\nu_e^TC^{-1}(1-\gamma_5)
\mu]$ etc.
The
effective strengths of these couplings arising from Fig. 2 (as well as
from Fig. 1 with $e$ replaced by a $\mu$) are found to be
\begin{eqnarray}
\epsilon_1^{e\mu} &=& {{\lambda_{123}\lambda'\_{311}} \over {2\sqrt{2}G_F
m_{\tilde{\tau}}^2m_{\tilde{\tau}^{c}}^2 }}
m_\tau(\mu {\rm tan}\beta+A_\tau m_0)
\nonumber \\
\epsilon_2^{e\mu} &=& {{\lambda'_{213}\lambda'_{131}} \over {2\sqrt{2}G_F
m_{\tilde{b}}^2m_{\tilde{b}^c}^2}} m_b (\mu {\rm tan}\beta + A_b m_0)
\end{eqnarray}
with similar expressions for $\epsilon_{1,2}^{\mu e}$.
The existing
limits on the $\lambda^{\prime}$ in Eq. (8) are $\lambda'_{213} \le 0.09,
\lambda'_{131} \le 0.26$, so that for the
squark masses of order 100 GeV, $\epsilon_2^{e\mu}$ can be as large as
$10^{-1}$ or so.  The branching ratio for
$\mu^-\rightarrow e^+$
conversion relative to $\mu$--capture can be obtained by a naive scaling
by the nucleon mass and we find
Br$(\mu^- \rightarrow e^+) \simeq 10^{-12}$ for this choice
of $\epsilon_2^{e\mu}$.  It is interesting that this is in the accessible
range of the current experiments.  The sensitivity of these
experiments is expected to improve by another order of magnitude in the
near future [16].

We note that the corresponding predictions for the left--right model is
down by several orders of magnitude.

In summary, we have discussed a new class of contributions to the
double lepton number violating processes which may arise in several
extensions of the standard model.  In particular, we find that
the existing experimental limits on neutrinoless double beta decay
lead to very stringent constraints on the
R--violating couplings in MSSM as well as the lepton number violating
Higgs couplings of the left--right symmetric model.  Furthermore, we find the
exciting possibility that for presently allowed range of parameters in the
MSSM, $\mu^-\rightarrow e^+$ conversion in nuclei is in the observable
range.  While we have focussed on only two classes of models, our
results are more general and should apply to other schemes, such as
$L$--violating leptoquark models.

\newpage

\section*{References}

\begin{enumerate}

\item W. C. Haxton and G. Stephenson, {\it Prog. in Part. and Nucl. Phys.}
{\bf 12}, 409 (1984); H. Grotz and H. Klapdor, {\it The Weak Interactions
in Nuclear, Particle and Astrophysics}, Adam Hilger,Bristol, (1990);
R. N. Mohapatra and P. B. Pal, {\it Massive Neutrinos in Physics and
Astrophysics}, World scientific, Singapore (1991); M. Moe and P. Vogel,
{\it Ann. Rev. Nucl. and Part. Sc.}, {\bf 44}, 247 (1994).

\item M. Doi, T. Kotani, H. Nishiura and E. Takasugi, {\it Prog. Theor.
Phys. Suppl}. {\bf 83}, 1 (1985).

\item A. Halprin, P. Minkowski, H. Primakoff and S.P. Rosen, {\it Phys. Rev.}
{\bf D13}, 2567 (1976).

\item R.N. Mohapatra, {\it Phys. Rev}. {\bf D34}, 909 (1986).

\item R.N. Mohapatra and J. Vergados, {\it Phys. Rev. Lett.}
{\bf 47}, 1713 (1981);
J. Schecter and J.W.F. Valle, {\it Phys. Rev.} {\bf D25}, 2951 (1982);
W.C. Haxton, S.P. Rosen and G.J. Stephenson, $ibid$., {\bf D26},1805 (1982);
L. Wolfenstein, $ibid$., {\bf D26}, 2507 (1982).

\item R. N. Mohapatra, {\it Phys. Rev.} {\bf D34}, 3457 (1986).

\item H. V. Klapdor-Kleingrothaus, {\it Prog. in Part. and Nucl. Phys.}
{\bf 32}, 261 (1994); A. Balysh et. al., {\it Phys. Lett.} {\bf B}
(to appear).

\item J. D. Vergados, {\it Phys. Lett.} {\bf B184}, 55 (1987);
M. Hirsch, H. V. Klapdor-Kleingrothaus and S. G. Kovalenko, Heidelberg
Preprint (1995).

\item L. Hall and M. Suzuki, {\it Nucl. Phys.} {\bf B231}, 419 (1984).

\item In addition, one can write a term $\mu_i L_iH_u$ in $W'$.  Although
this term
can be rotated away by a field redefinition, the VEVs of
the scalar neutrinos cannot be set to zero in general then.  In supergravity
models, however, such VEVs get induced only by the effect of large
Yukawa couplings, in which case the sneutrino VEVS will remain zero.

\item There is also the possibility of setting $\lambda=\lambda'=0$, with
$\lambda^{\prime\prime} \ne 0$, in which case $B$ is violated, but $L$ is not.
We shall not pursue this alternative here.

\item V. Barger, G. Giudice and T. Han, {\it Phys. Rev.} {\bf D40},
2987 (1989).

\item R.N. Mohapatra and G. Senjanovi\'c , {\it Phys. Rev. Lett}.
{\bf  44}, 912 (1980); {\it Phys. Rev}. {\bf D23}, 165 (1981).

\item Note that in the limit $\kappa' = 0$, this coupling does not
induce a $\langle \Delta_L \rangle$.  See K.S. Babu and R.N. Mohapatra,
{\it Phys. Rev. Lett}. {\bf 64}, 9 (1990).

\item M. Schwarz, {\it Phys. Rev.} {\bf D40}, 1521 (1989).

\item W. Bertl, private communication.

\end{enumerate}

\noindent{\bf Figure Caption:}

\vspace{4mm}

\noindent{\bf Fig.1:} The dominant diagram contributing to $\epsilon_2^{ee}$
type Four-Fermi term (Eq. (1)) in the supersymmetric model.

\vspace{4mm}

\noindent{\bf Fig.2:} The diagram contributing to $\epsilon_1^{e\mu}$ type
Four-Fermi terms relevant for $\mu^- \rightarrow e^+$ conversion in the
supersymmetric model.

\vspace{4mm}

\noindent{\bf Fig.3:} The vector--scalar exchange diagram for
$\beta\beta_{0\nu}$ in the left--right symmetric model.

\end{document}